# Surface terraces in pure tungsten formed by high-temperature oxidation

Hongbing Yu[1*], Suchandrima Das[1], Junliang Liu[2], Jason Hess[3], Felix Hofmann[1†]

(1) Department of Engineering Science, University of Oxford, Parks Road, Oxford, OX1 3PJ, UK

(2) Department of Materials, University of Oxford, Parks Road, Oxford, OX1 3PH, UK

(3) Materials Science & Scientific Computing, UKAEA, Abingdon, OX14 3DB

*hongbing.yu@eng.ox.ac.uk*   †*felix.hofmann@eng.ox.ac.uk*

**Abstract**: We observe large-scale surface terraces in tungsten oxidised at high temperature and in high vacuum. Their formation is highly dependent on crystal orientation, with only {111} grains showing prominent terraces. Terrace facets are aligned with {100} crystallographic planes, leading to an increase in total surface energy, making a diffusion-driven formation mechanism unlikely. Instead we hypothesize that preferential oxidation of {100} crystal planes controls terrace formation. Grain height profiles after oxidation and the morphology of samples heat treated with limited oxygen supply are consistent with this hypothesis. Our observations have important implications for the use of tungsten in extreme environments.

**Main text:**

Tungsten (W) is scientifically interesting as it has the highest melting and boiling temperatures and the highest yield strength among the un-alloyed elements. Additional beneficial properties of high thermal and electric conductivity [1] make it the front-runner material for many high-temperature applications. Some examples include light bulbs, rocket engine nozzles, accelerator targets, and electron microscopy gun filaments and tips [2,3]. Recently, tungsten has also captured the interest of the nuclear fusion industry, where it is being considered as candidate for plasma-facing armour components [4–7].

In addition, as a prototypical bcc material, tungsten has been extensively used to study surface morphology and evolution [8–11]. In particular, surface terraces and steps are widely investigated since surface defects like steps, ledges, kinks, vacancies and individual adatoms are strong adsorption sites [9,12]. Clean and atomically smooth terraces aligned to a perfect crystal orientation can be readily prepared on the surface of tungsten by high temperature annealing (1500 °C) under ultrahigh vacuum conditions ($1.3\times10^{-10}$ mbar) [9,11]. The steps separating these terraces are normally one atomic layer high, and as such termed mono-atomic steps. However, with a controlled supply of oxygen during high temperature heat treatment, multi-atomic steps can be formed [11]. The orientation of the facets tends to be the nearest low-index plane. Therefore, to obtain multi-atomic steps, a well-controlled initial surface orientation is required with little angular deviation from the desired terrace orientation [9,11].

Surprisingly, after heat treatment of polycrystalline tungsten at 1500 °C in a vacuum furnace at pressures of an order of $10^{-5}$ mbar, we observed the emergence of large surface terraces. The surface morphology shows a striking orientation dependence. Grains with surface orientation close to {111} show a stacking of cubic corners or/and large-scale step terraces with facets aligned to {100} planes, while grains with {100} surface have a flat and smooth surface. This is remarkable particularly since {100} planes have higher surface energy than {111} planes [13]. Moreover, the steps we observe are hundreds of nanometres high. In this paper, we explore the underlying mechanism for this terrace structure formation.

High purity polycrystalline tungsten (> 99.97%) sheet (1 mm thick), purchased from Plansee, was used for this study. The as-rolled sheet was cut into $10\times10\times1$ mm$^3$ pieces. The surface of these samples was ground using 600 to 1200 grade abrasive papers, followed by diamond paste polishing and a final electropolishing step to prepare a flat surface prior to heat treatment. Electropolishing was done in a bath of 1% NaOH in deionized water, at a voltage of 8 V at room temperature for 1 to 2 minutes. The samples were then heat-treated at 1500 °C for 24 h in a vacuum furnace with pressure maintained at ~$10^{-5}$ mbar.

The surface morphology after annealing was characterized in Zeiss Merlin and Crossbeam 540 scanning electron microscopes (SEM). Electron backscattering diffraction (EBSD) was used to determine the orientation of grains and energy dispersive X-ray spectroscopy (EDX) was used to probe the content of oxygen in the surface layer. EBSD was performed at 20 keV, 15 nA current, and EDX was conducted using 5 keV, 3 nA current. Atomic force microscopy (AFM) and SEM stereo-imaging were employed to characterise the height profile of the surface.

Figure 1 shows the surface morphology of the pure polycrystalline tungsten after heat treatment at 1500 °C for 24 h. Terrace-like structures are observed in some of the grains (Fig. 1(a) and (c-d)). The orientations of the grains are depicted by the inverse pole figure (IPF) map shown in Fig.1(b). A comparison of Fig. 1 (a) and (b) shows that there is a correlation between the surface structure and the surface orientations. Grains with surface orientation close to {111} crystallographic plane (blue in the IPF map) all have a terrace structure on the surface, while grains with surface orientation close to {100} orientation (red in the IPF map) are generally smooth and flat. As the grain orientation deviates from the {111} orientation, the surface structure become less pronounced, as demonstrated by the purple grain in subset 2 ( Fig. 1 (d)), which only shows very shallow terraces. More examples of the orientation dependence of surface structure can be found in supplementary Fig. S1 and S2.

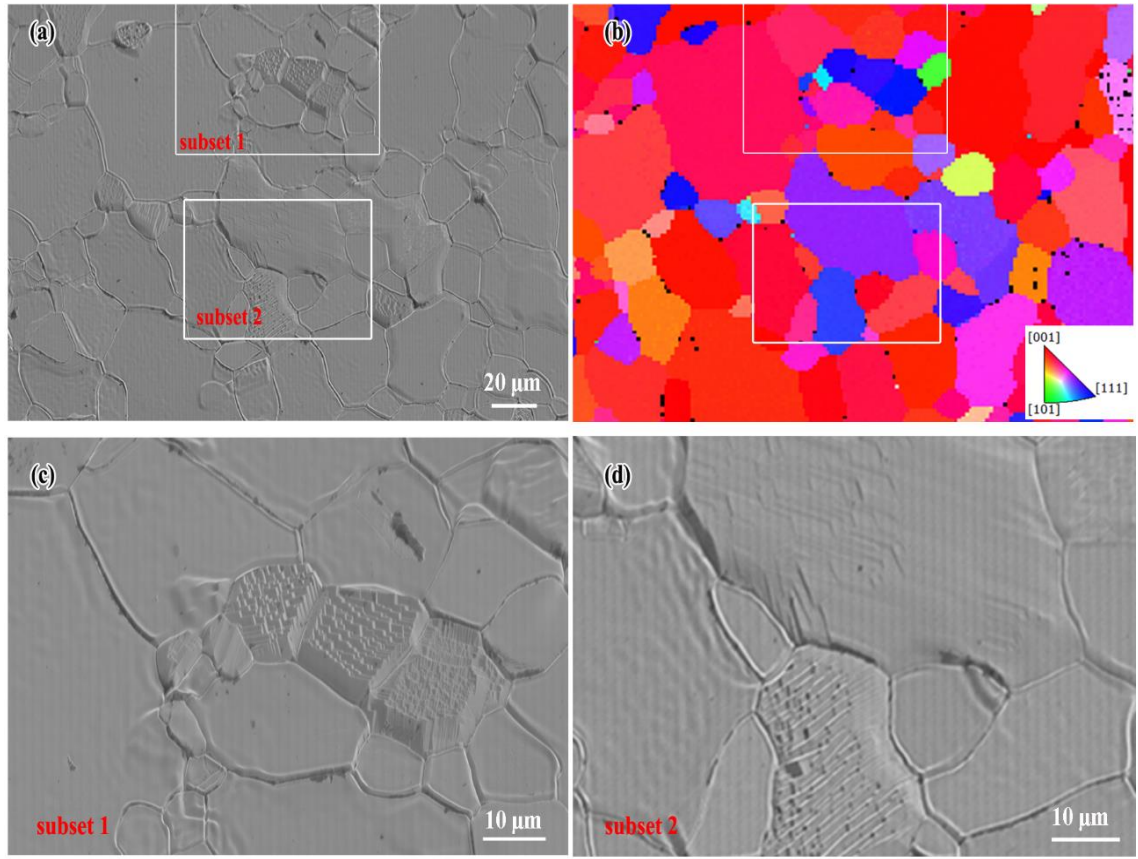

Fig. 1 Orientation dependence of the surface morphology of a pure polycrystalline tungsten sample with initially flat surface after heat treatment at 1500 °C for 24 h. (a) Secondary electron image, (b) Inverse pole figure colour map of out-of-plane crystal orientation. (c) and (d) zoomed in images for the subsets shown (a).

An interesting question arises regarding the alignment of the terrace facets with respect to crystallographic orientation. To address this, a SEM image was recorded at 0° tilt from a region also studied by EBSD. Figure 2 (a) shows the surface terraces in three grains viewed at 70° tilt. The corresponding EBSD map in Fig. 2(b) shows that all three grains have close to {111} surface orientation (misorientations range from 4° to 8°). When viewed at 0° tilt, the edges of the terraces in all three grains are parallel to one of the <100> crystallographic directions (Fig. 2 (c) and (d)). This means that the facets of the terraces must be {100} crystallographic planes. The {100} facets are also found to dominate in grains that deviate 15° to 30° from the {111} orientation, though a small portion of ledges are not {100} planes (supplementary Fig. S1). The height of steps can be estimated from Fig.2 (a) and (c) using triangulation. For example, the height of the step marked in Fig. 2 (c) and (d) (pink arrow) is 700 nm along the [001] direction. This is orders of magnitude larger than step heights previously reported in the literature, even in the presence of oxygen [11].

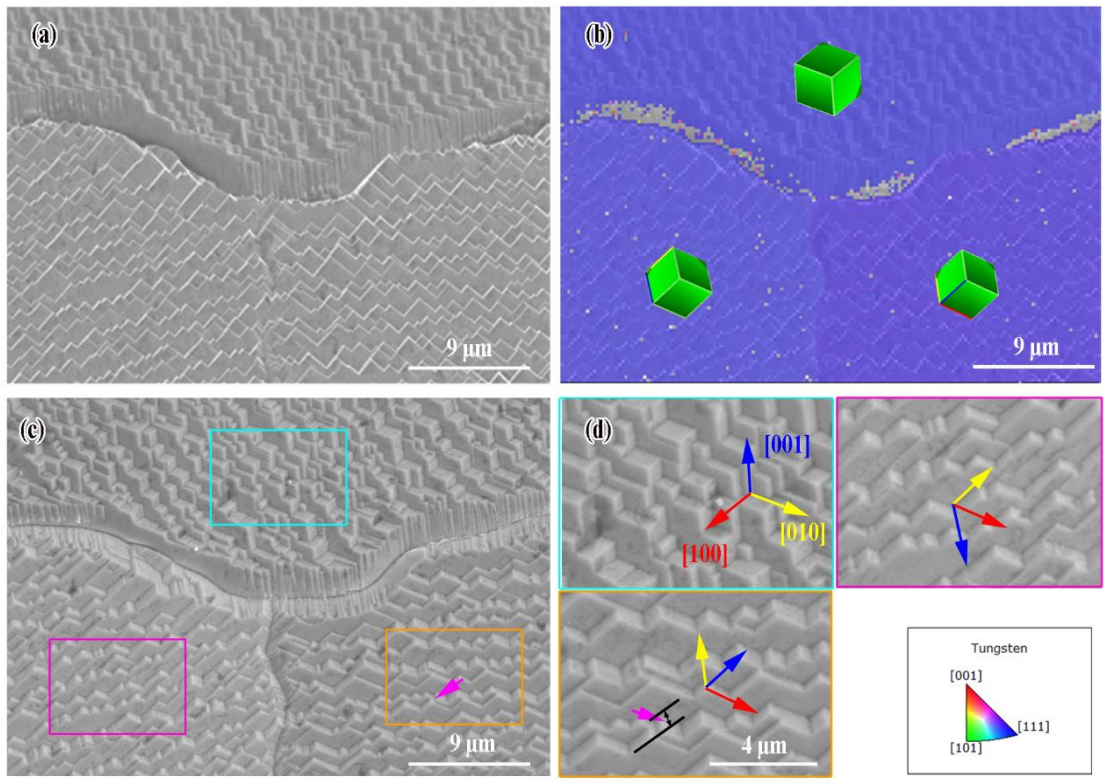

Fig. 2 Surface orientation of the terrace facets. (a) Secondary electron imaging of the terraces at 70° tilt, (b) Inverse pole figure colour map of out-of-plane orientation, (c) secondary electron image of the same region with 0° tilt, (d) the zoomed in regions from (c). The green cubes in (b) show the crystallographic orientation of the corresponding grains. The pink arrow in (c) and (d) points the step of which the height was estimated. Red, yellow and blue arrows in (d) point in the directions of [100], [010] and [001] axes respectively.

The presence of terraces with pre-dominantly {001} facets is surprising. Though reports of the absolute surface energy of tungsten show some variation [13,14], the (100) plane is generally believed to have the highest surface energy followed by the (111) and finally the (110) plane (with the lowest surface energy). Our observation means that after thermal treatment a {111} surface with lower surface energy turns into terraces with higher surface energy facets, resulting in a total surface energy increase of more than 70% for perfectly oriented {111} grains. Together with the observed step height of hundreds of nanometres and the large deviation of facets from the original surface orientation, this makes it unlikely that thermal diffusion of surface atoms, driven by surface energy minimisation, causes the surface terraces. The surface structure may also be affected by recrystallization-induced roughening. To probe this effect, an as-rolled and a fully annealed sample were polished to a mirror finish and then heat treated at 1500 °C for 10 h. SEM images show little difference in the surface structures on these samples (see supplementary Fig. S3), suggesting that recrystallization roughening does not play an important role.

A possible mechanism could be the orientation dependence of oxidation rate. Although tungsten has the highest melting point, it readily oxidises above 400 °C [15–17]. Tungsten trioxide ($WO_3$), along with its variants like $W_{18}O_{49}$ and $WO_{2.9}$, is the main oxide formed during oxidation up to temperatures of 2000 °C [16–18]. Once formed, the oxide readily sublimes above 1000 °C [19,20]. A recent study showed a strong surface orientation dependence of the oxidation rate of pure W at temperatures between 450 °C and 600 °C [21]. Grains with {001} surface have the highest oxidation rate; more than double of {111} oriented grains. The oxidation rate of grains with {110} surface orientation lies between these two extremes. At temperatures above 1000 °C the surface oxide sublimes once it forms [15]. In this case, the oxidation rate is controlled by the rate of oxygen supply and the probability of oxygen reacting with the tungsten surface, which also depends on orientation. A previous study, conducted at 2050 °C under $2 \times 10^{-4}$ mbar partial pressure of oxygen, showed a 5 to 6 times higher probability of oxygen reacting with the {001} oriented tungsten surface as compared to the {111} and {110} surfaces [22]. Hence, we hypothesise that under the experimental conditions used in this study, initially grains with {111} surfaces are preferentially oxidised along {100} planes. The final observed terrace structures result from the exposure of the un-oxidized {100} facets following sublimation of the oxide layer.

To test this hypothesis of oxidation-driven restructuring, we performed two further experiments. Firstly, we measured a mass loss of 0.078 g after heat treatment of a 500 mm$^2$ W plate (1500 °C for 20 h, same vacuum conditions as previous samples), due to the sublimation of tungsten oxide, corresponding to a thickness reduction of ~4 μm for each side. Secondly, two identical samples, one sealed in a ϕ 6mm× 80 mm high vacuum quartz tube (termed QT sample), and one not enclosed (termed furnace sample), were held at 1200 °C for 24 h in the vacuum furnace. The QT sample had very limited oxygen supply as the quartz tube serves as a gas barrier even at high temperatures. The furnace sample, on the other hand, is continually exposed to a low partial pressure of oxygen (on the order of $10^{-6}$ mbar) due to a small amount of gas leakage from joints and connections of the vacuum furnace.

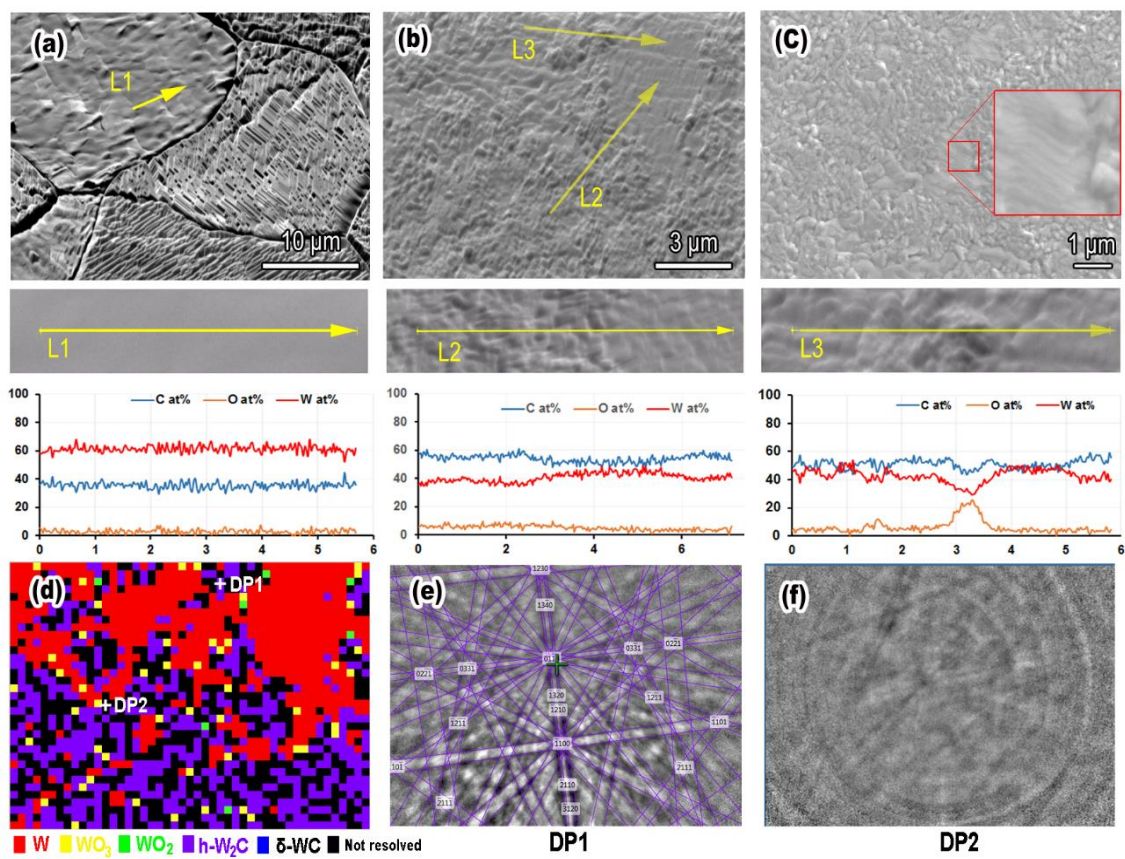

Fig. 3 Surface morphology after heat treatment at 1200 °C for 24 h. (a) Sample outside the quartz tube in the vacuum furnace, (b) and (c) sample heat treated inside a sealed quartz tube in the vacuum furnace. A lower magnification image of the quartz tube sample can be seen in supplementary Fig. S5. The line profiles show the elemental distribution along the corresponding line scans determined by EDX. (d) EBSD phase map acquired from (b). (e) and (f) show the diffraction patterns at the points indicated by the crosses in (d).

Fig. 3 shows the surface morphology of the furnace sample and QT sample after heat treatment. Large-scale terraces are evident on the surface of some grains in the furnace sample (Fig. 3(a)), similar to the sample heat treated at 1500 °C (Figs. 1 and 2). A completely different morphology is observed on the QT sample. Dense irregular and particle-like features dominate the surface. A few relatively flat, smooth surface patches, some of which contain shallow terraces, are also visible (Fig. 3(b) and (c)). EDX line scans on the furnace sample show little residual oxygen on the surface (3 at % on average). However, a high content of carbon is found on the furnace sample surface (35 at% on average), presumably due to carbon contamination after heat treatment, since this carbon was readily removed by plasma-cleaning (supplementary Fig. S6). The L2 line scan on the QT sample covers the transition region, from an irregular, rough area to a smooth patch. The carbon content in the QT sample is significantly higher than the furnace sample, between 50-60%. There is no significant difference in the carbon content between the rough area and flat patch, but higher oxygen concentration (6.5 at%) in the rough area than the flat patch (3.6 at%, similar to the furnace sample). In some particularly large lumps on the rough surface we found oxygen concentrations of up to 30 at% (e.g. L3 in Fig. 3(c)). EBSD of the region in Fig. 3(b) showed that the flat patch gave sharp diffraction patterns that could be reliably indexed as tungsten (Fig. 3(d)). On the other hand, the irregular rough area is composed of mainly hexagonal $W_2C$ ($P\bar{3}m1$, a = 0.3 nm, c = 0.47 nm) and other unindexed regions. Typical EBSD patterns of the hexagonal $W_2C$ and the unindexed region are shown in Fig. 3 (e) and (f), respectively. $WO_2$ ($P2_1/c$), $WO_3$ (9 different crystal structures) and $\delta$-WC ($P\bar{6}m2$, a = 0.2906, c = 0.2837) were also included in the phase list for indexing. Interestingly the EBSD pattern from the unindexed region shows rings (Fig. 3 (f)), which together with a spike in oxygen concentration (L3 line scan) suggests the presence of residual oxide on surface. However, the measurements do not reveal the type of oxide formed.

The irregular rough surface and the residual W oxides and $W_2C$ in the QT sample can be attributed to two factors. Firstly, as the encapsulated sample cools, the volatile $WO_3$ may redeposit on the sample surface in absence of an exit route from the quartz tube. For the furnace sample $WO_3$ vapour is continually removed by the vacuum pump, leaving clean surfaces. Secondly, as the oxygen in the quartz tube is used up, $WO_2$ as well as W carbide become stable [16]. However, due to the very limited supply of carbon, lower carbide hexagonal $W_2C$ forms instead of the more stable $\delta$-WC. Though $W_2C$ is known to decompose into $\delta$-WC and tungsten below 1250 °C [23], our results show that a thin layer of $W_2C$ may be retained on the tungsten surface even at room temperature. A possible explanation for this may be that the defects required for the phase transformation are lost at surface features and grain boundaries, which act as strong sinks [24].

Comparison of the furnace and QT samples suggests that oxidation plays an important role in the formation and coarsening of the surface terraces. If the preferential oxidation of {100} surfaces is indeed the dominant mechanism, one would expect grains with surface orientation close to {111} to stand proud of the surface, compared to grains with {100} orientation. This is confirmed by AFM measured surface height profiles and SEM stereo-imaging of the 1500 °C annealed sample (Fig. 4). The AFM map shows that grains with terraces are 200 nm higher than the surrounding flat grains. This result is consistent with SEM stereo-imaging where we found grains with terraces that were up to 1 μm higher than neighbouring flat grains.

These results have important implications for the use of tungsten in future fusion reactors. The base pressure in fusion reactor vacuum vessel will be as low as $10^{-9}$ mbar [25,26]. However, the oxidation of tungsten is a major safety concern since large quantities of volatile tungsten oxide may form and sublime in a loss-of-coolant accident with simultaneous leakage of the vacuum vessel. This radioactive and volatile oxide would pose a health risk if released into the atmosphere. In response, researchers have tried to develop oxidation resistant tungsten alloys by adding alloying elements such as Cr and Y, that could form a protective oxidation layer on the surface [20,27–29]. Our study shows that texture control could be another strategy to address this challenge. Importantly a preferred (111) texture is compatible with recommendations from other studies considering the effects of surface orientation on sputtering rate [30] and helium-induced surface nanofuzz [31]. That is, the (111) orientation has the lowest sputtering rate, while the {001} surface should be avoided as it is most susceptible to nanofuzz formation.

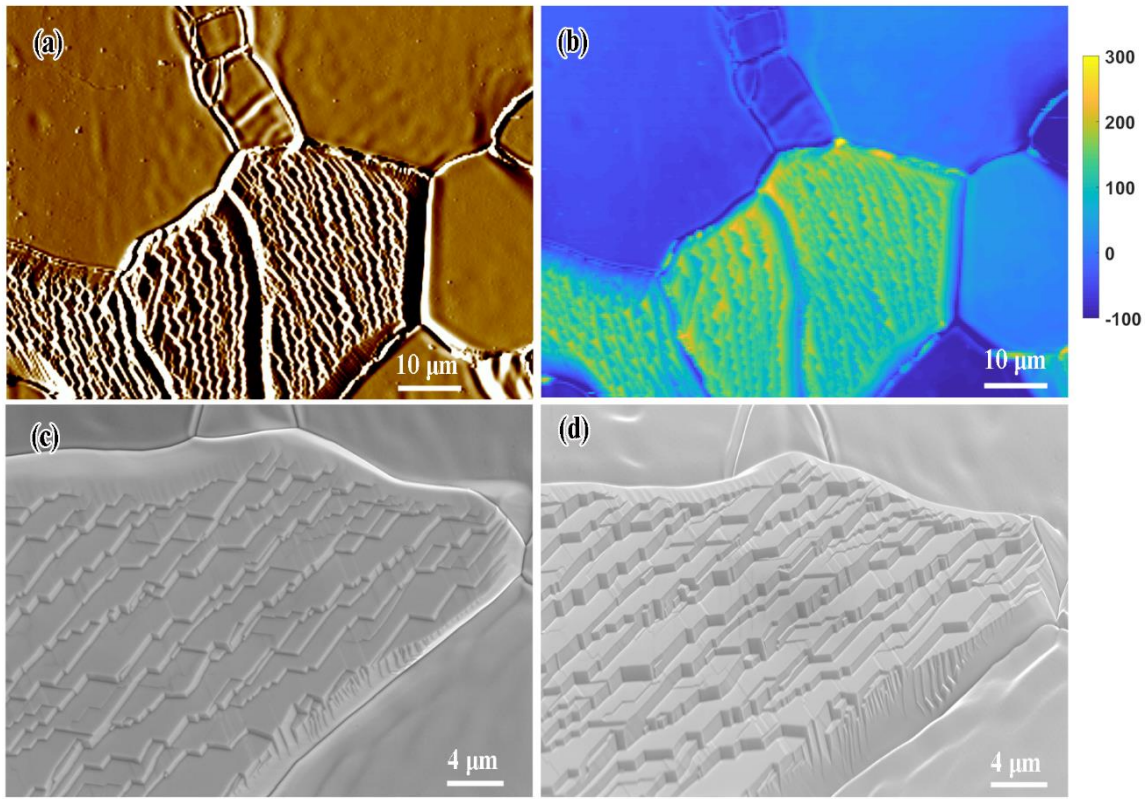

Fig.4 Microstructure and height profiles. (a) AFM gradient image, (b) AFM height image. (c) and (d) SEM images taken at 0° tilt and 70° tilt with tilt correction applied.

In a summary, we have observed the formation of surface terraces with {001} facets in pure tungsten grains with surface orientation close to the {111} plane after annealing at 1500 °C in high vacuum. Our analysis shows that preferential oxidation of {001} planes, rather than diffusion of surface atoms, is the main mechanism driving the formation of this interesting structure. The grains with terraces are observed to be at a higher elevation than surrounding grains with smooth surfaces. This suggests a lower oxidation rate of {111}-oriented grains. In addition, high temperature phase $W_2C$ was observed to be retained on the surface of tungsten even at room temperature.

**Acknowledgements**

This work was funded by Leverhulme Trust Research Project Grant RPG-2016-190. The authors acknowledge use of characterisation facilities within the David Cockayne Centre for Electron Microscopy, Department of Materials, University of Oxford and within the LIMA Lab, Department of Engineering Science, University of Oxford.

# Supplementary material

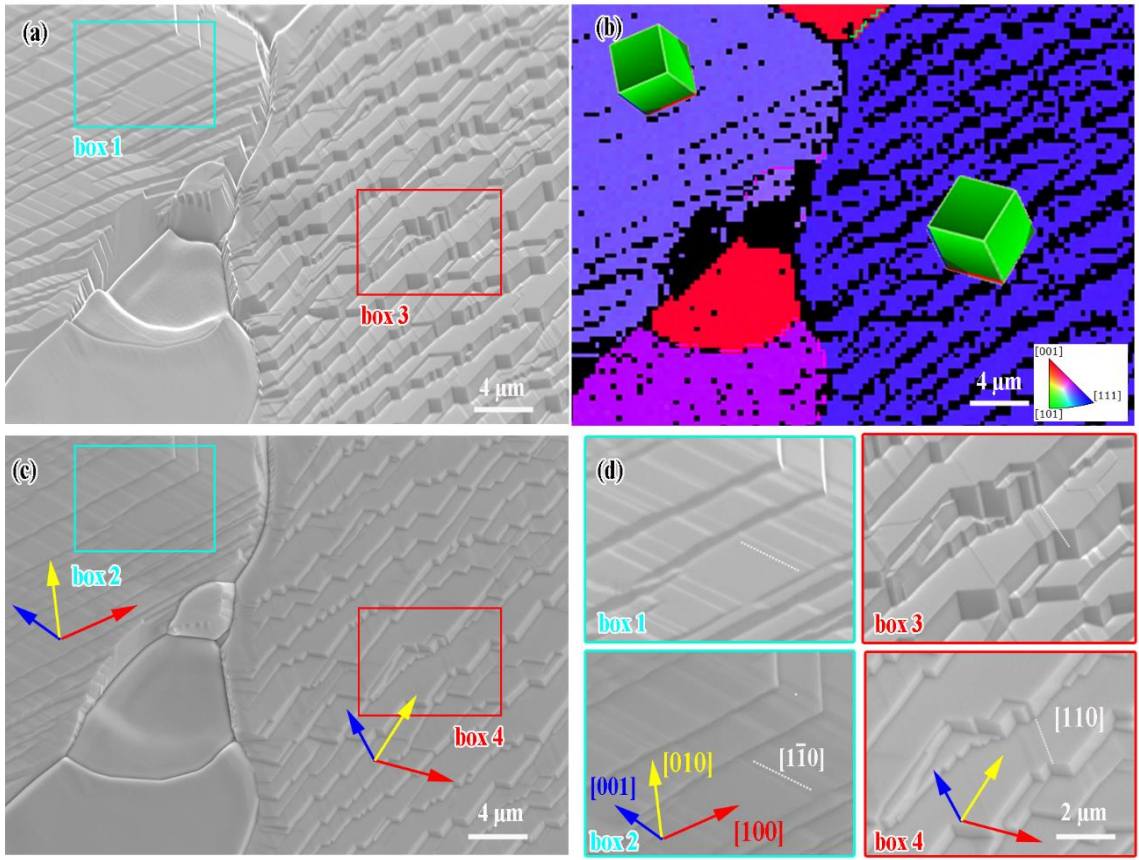

Fig. S1 Surface morphology of two grains that deviate from {111} surface orientation. The surface orientation in the left grain deviated ~ 30° from the {111} plane, while the right one deviated ~ 15°. The majority of the facets are found to be {100} planes. Ledges are aligned with their edges parallel to [100] directions. (a) Secondary electron imaging of the terraces at 70° tilt, (b) Inverse pole figure colour map of out-of-plane orientation, (c) secondary electron image of the same region with 0° tilt, (d) the zoomed in regions from (c). The green cubes in (b) show the crystallographic orientation of the corresponding grains. Red, yellow and blue arrows in (d) point in the directions of [100], [010] and [001] axes respectively.

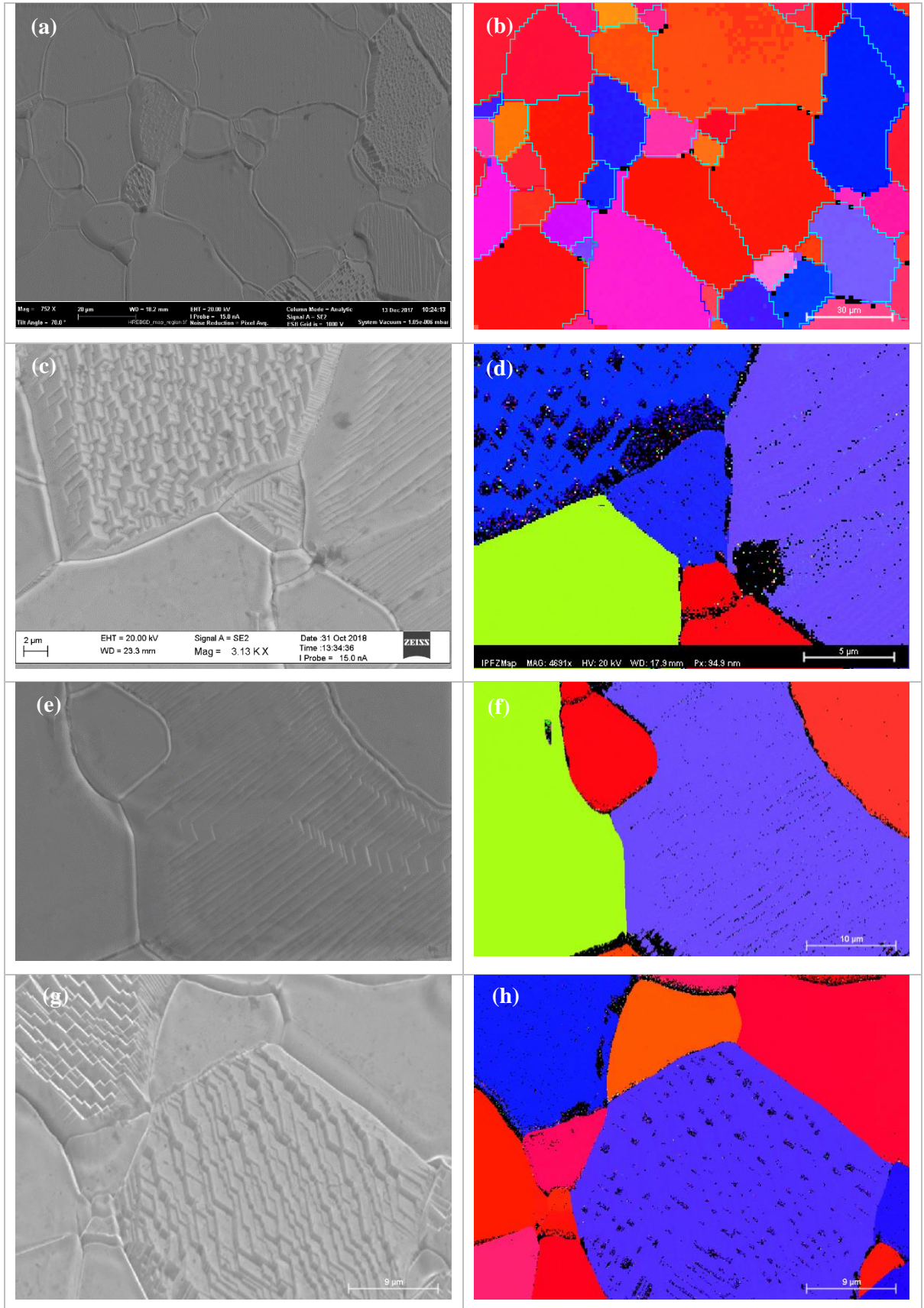

Fig. S2 The correlation between the surface morphology and the grain orientation from randomly selected regions. Blue grains which have {111} surface orientation show terraces structure on the surface, while green and red grains which are {110} and {001} oriented do not show terraces.

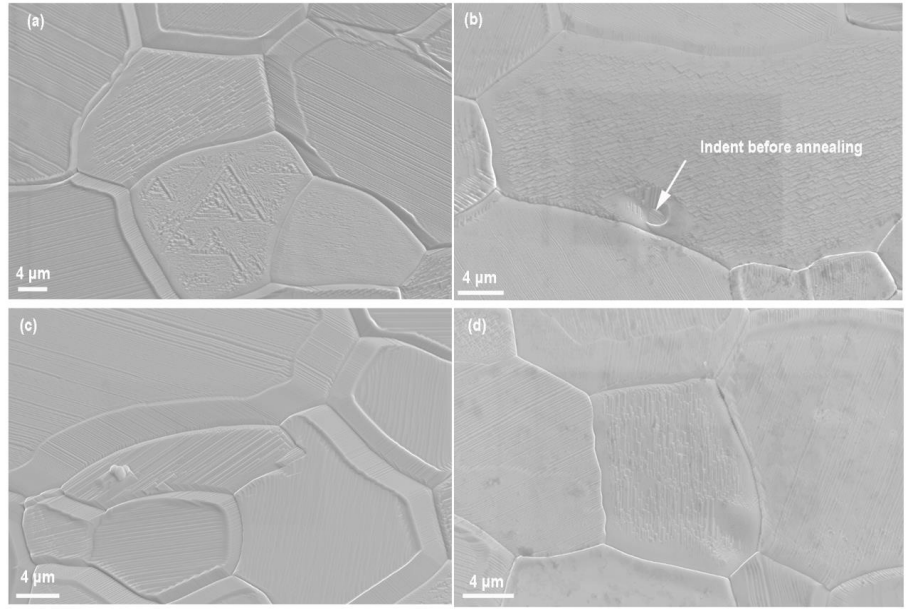

Fig. S3 Surface structure after annealing at 1500 °C for 10 h from mirror polished samples. (a) and (c) from fully recrystallized W, (b) and (d) from as-rolled W. Both samples show surface terraces. This suggests that recrystallization-induced roughening does not play an important role in the formation of surface terraces.

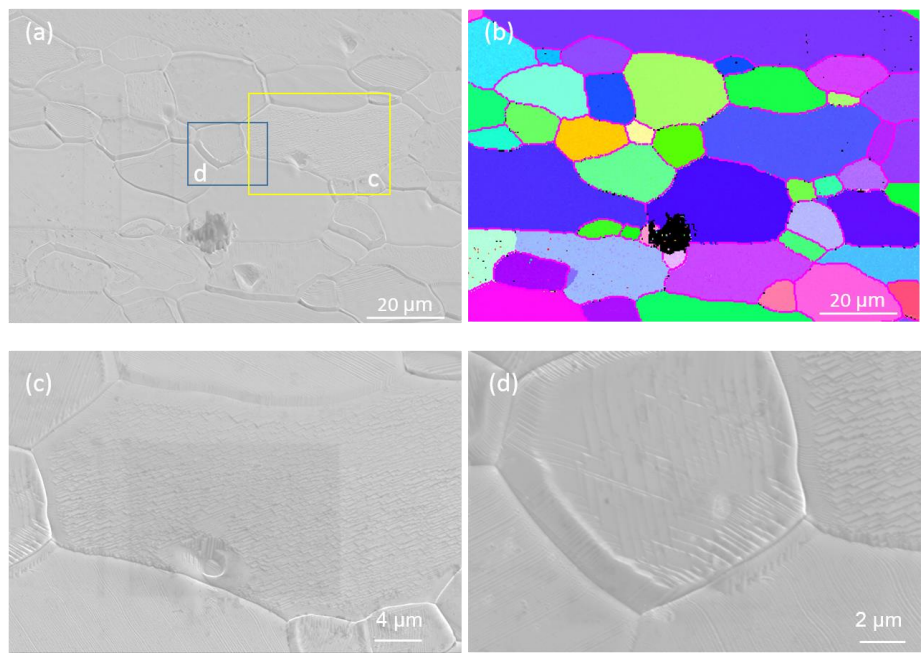

Fig. S4 The correlation between the surface structure and surface orientation in an as-rolled sample after annealing at 1500 °C for 10 h. (a), (c) and(d) Secondary election imaging, (b) IPF along the out of plane direction. Green grains shows steps on the surface after annealing at 1500 °C for 10 h.

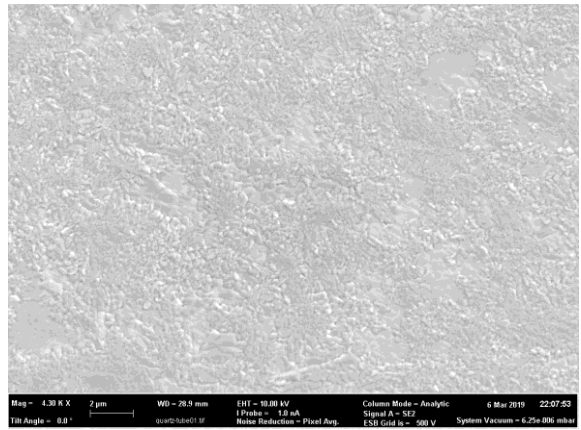

Fig. S5 Microstructure of the quartz tube sample at lower magnifications.

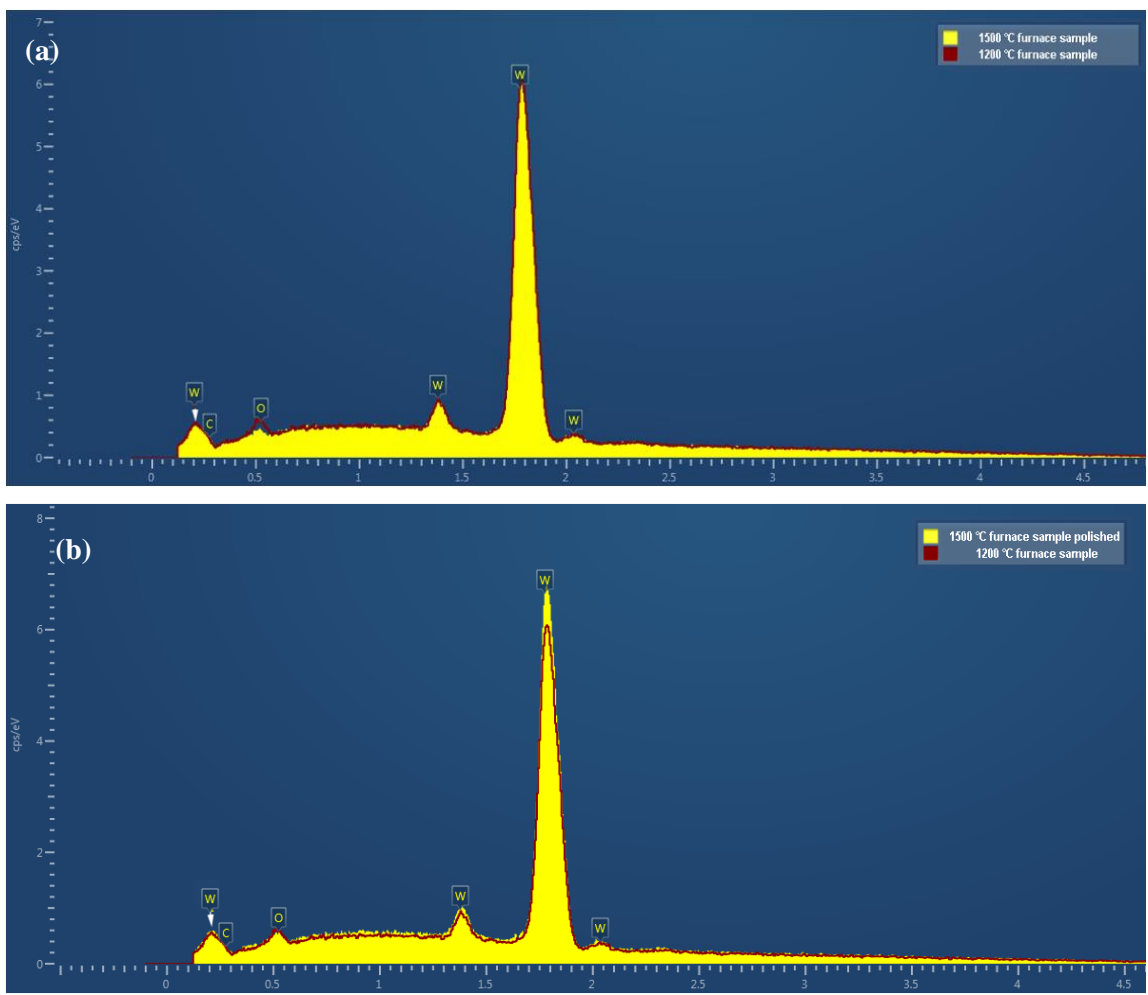

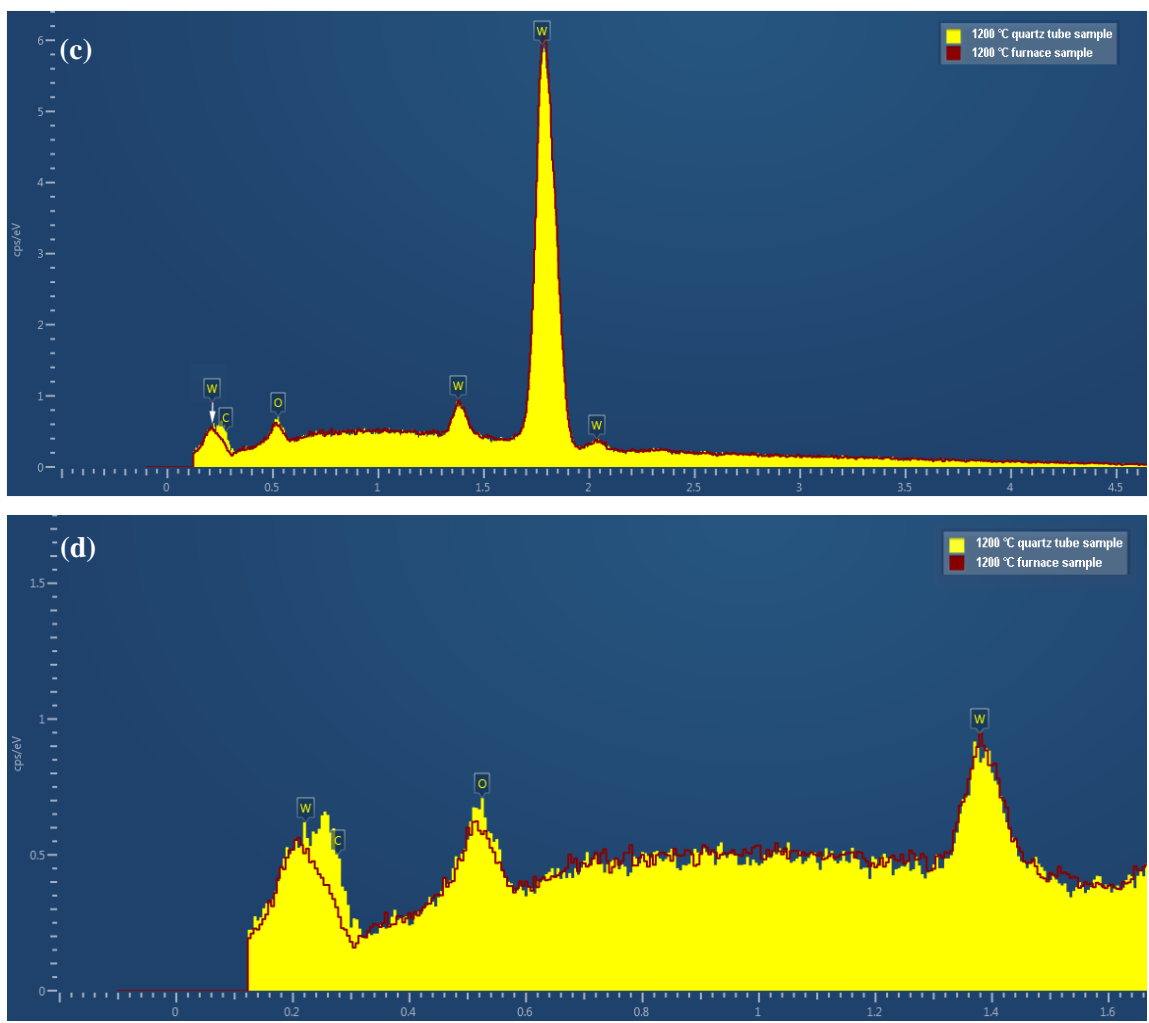

Fig. S6 The comparison of the EDX spectrum of 1200 °C furnace sample (after 10 min plasma cleaning) and other samples. (a) 1200 °C furnace sample vs 1500 °C furnace sample, (b) 1200 °C furnace sample vs 1500 °C furnace sample (after electropolishing), (c) and (d) 1200 °C furnace sample vs 1200 °C quartz tube sample (after 10 min plasma cleaning). There is discernible difference between the plasma cleaned 1200 °C furnace sample, the 1500 °C furnace sample and the electropolished sample after 1500 °C heat treatment. There is no obvious carbon detect in any of these three samples. However, a significant carbon peak is observed in the 1200 °C quartz tube sample, which is in good agreement with W$_2$C on the surface probed by EBSD.